\documentclass[prl,superscriptaddress,twocolumn,showpacs,preprintnumbers,amsmath,aps]{revtex4}
\usepackage{amsmath,amssymb}
\usepackage{epsfig,graphicx,hyperref}
\usepackage{dcolumn}
\usepackage{bm}

\begin{document}
\title{Kibble-Zurek mechanism and infinitely slow annealing through critical points}
\author{Giulio Biroli}
\affiliation{Institut de Physique Th\'eorique, CEA, IPhT, F-91191 Gif-sur-Yvette, France and CNRS, URA 2306}
\author{Leticia F. Cugliandolo}
\affiliation{Universit\'e Pierre et Marie Curie -- Paris VI, LPTHE UMR 7589,
4 Place Jussieu,  75252 Paris Cedex 05, France}
\author{Alberto Sicilia}
\affiliation{Department of Chemistry, University of Cambridge, Lensfield Road, CB2 1EW, Cambridge, UK}

\date{\today}

\begin{abstract}
We revisit the Kibble-Zurek mechanism by analyzing the dynamics of phase ordering systems 
during an infinitely slow annealing across a second order phase transition. 
We elucidate the time and cooling rate dependence of the typical growing length and 
we use it to predict the number of topological defects left over in the symmetry broken phase as a function of time, 
both close and far from the critical region. Our results extend the Kibble-Zurek mechanism and reveal its limitations. 
\end{abstract}

\pacs{}

\maketitle

The out of equilibrium dynamics induced by a quench are
the focus of intense research~\cite{Alan,Cugliandoloreview}.
Interesting realizations are quenches through a second order phase
transition, which take the system from the symmetric 
phase into the symmetry broken one. Below the
transition, times scaling with the system size are needed to reach
equilibrium and to realize the spontaneous symmetry breaking
process.  Before this--typically unreachable--asymptotic limit the
symmetry is broken only locally: the system is formed by ordered
regions of size growing with time~\cite{footnote0}. Only when 
this size reaches the order of the volume of the sample the
symmetry is broken globally and the spatial average of the order 
parameter deviates from zero.
The majority of theoretical studies focused on the dynamics after
infinitely rapid quenches although experimentally quenches are
performed at finite speed. Indeed,
since the typical time-scale on which the system evolves is its age,
i.e. the time elapsed since crossing the critical point, finite quench
time-scales ($\tau_Q$) eventually become short compared to the
relaxation time. Thus, they alter the out of equilibrium dynamics at short
times only.
The opposite limit of an extremely slow annealing, corresponding to very
long $\tau_Q$, needs a separate treatment.  Surprisingly, this has not
been studied in detail in the statistical physics literature with,
however, some  exceptions for disordered 
systems~\cite{fisherhuse,yoshino,leticiajp}. It has, instead, attracted a lot of attention
within the cosmology and, more recently, the condensed matter
communities. An explanation of the slow dynamics induced by this
protocol was given by the so-called Kibble-Zurek (KZ)
mechanism~\cite{original-Kibble,original-Zurek,report-Zurek,LesHouches}. This
is an {\it equilibrium scaling argument} that yields an estimate for the
density of topological defects left over in the ordered phase as a
function of the quenching rate close to the critical point.  The argument has
been recently generalized to study very slow `quantum annealing' across a
quantum phase transitions in isolated systems~\cite{Zurek-Zoller,Polkovnikov,Dziarmaga}.

The aim of this work is to obtain a more complete picture of the slow
dynamics induced by an extremely slow annealing.  With numerical and
analytical arguments we unveil the limitations of the KZ approach and
we obtain a full scaling description of the slow dynamics. Our main
result is that the dynamic evolution is characterized by a first
adiabatic regime in agreement with KZ, followed by critical coarsening
and, finally, standard coarsening at very long times.
We find a {\it new} universal scaling function that
characterizes the growth of the correlation length out of equilibrium
under slow cooling procedures and we relate it to the number of 
topological defects in cases in which these exist.

We start our discussion by recalling the KZ
mechanism~\cite{original-Zurek,report-Zurek,LesHouches}. We take  
a system in  equilibrium at equilibrium at a value $g_0> g_c$ of the 
control parameter in the symmetric phase 
and subsequently anneal it at finite rate (in typical situations
$g$ corresponds to temperature).  As
KZ, we focus on the protocol $g(t) = g_c(1-t/\tau_Q)$ starting
from, say, $g_0=g(-\tau_Q)=2g_c$. Henceforth we use the standard
notation of dynamical critical phenomena~\cite{Hohenberg-Halperin} and
we set the microsocopic time and length scales to one.  Far from the
critical point the equilibrium relaxation time, $\tau_{eq}$, is barely
larger than the microscopic time. Thus, for very small annealing rate,
i.e. very long $\tau_Q$, the system evolves adiabatically and remains
in equilibrium at the running $g(t)$.  However, this regime must
inevitably break down since $\tau_{eq}$ diverges at the critical point as
 $|\Delta g|^{-\nu z_{eq}}$ with $\Delta g \equiv g-g_c$. KZ argued
that the end of the adiabatic regime occurs when the remaining time,
$\hat t$, needed to reach $g_c$ becomes smaller than $\tau_{eq}$. This
is certainly a lower bound and yields  $\hat t
\propto\tau_{eq}(\hat g) \propto \tau_Q^{\nu z_{eq}/(1+\nu z_{eq})}$
with $\hat g=g(-\hat t)$. The distance from the critical
point at $-\hat t$ is $\Delta\hat g\propto \tau_Q^{-1/(1+\nu
  z_{eq})}$.  
KZ assumed that after $-\hat t$ the topological
defect configuration remains
frozen, in the sense that the order parameter ceases to evolve. 
In this so called `impulse' regime the effect of
lowering $g$ is to reduce fluctuations, which are in
general of thermal origin since very often $g$ is related to the
temperature.  The main prediction of KZ is the number of topological
defects, $N$,  at the symmetric instant $\hat t$ where the coupling constant
equals $g_c-\Delta\hat g$.  Within their approach $N$ is inherited
from the configuration at $-\hat t$, it is therefore equal to the
number of defects in equilibrium at $\hat g$, and it is estimated
to be $N(\hat t) \simeq [f^2 \xi_{eq}(\hat g)]^{-d}\simeq f^{-2d}
|{\Delta \hat g}|^{d\nu}$ with $f$ of the order of one.  Knowing the
$\tau_Q$-dependence of $\Delta\hat g$ allows one to derive the
$\tau_Q$-dependence of $N$:
\begin{equation}
 N(\hat t)\propto \tau_Q^{-d\nu/(1+\nu z_{eq})}
\;\;\;\; \mbox{at} \;\;\;\; 
\hat t \propto \tau_Q^{\nu z_{eq}/(1+\nu z_{eq})}
\; . 
\label{eq:KZ-prediction}
\end{equation} 
We stress that for each $\tau_Q$ this expression should be measured at
the special instant $\hat t(\tau_Q)$ after the phase transition.
The system's behaviour at $t>\hat t$ is not fully addressed by KZ. 
In some publications it is assumed that any further
evolution~\cite{Laguna-Zurek,Yates-Zurek} can be neglected, whereas
in others it is reckoned that after
$\hat t$ the system resumes its out of equilibrium evolution with a
mechanism that depends on the problem at hand (domain growth,
vortex-anti-vortex diffusion and annihilation, etc.)~\cite{report-Zurek} and the
density of defects may therefore continue to decrease although no detailed study
was performed. 
Numerous numerical~\cite{Laguna-Zurek, Yates-Zurek, Bettencourt-Zurek,
  Stephens,Suzuki} and experimental~\cite{no,yes,kibble-exp,newexp,kirtley} papers tested the
quantitative consequences of the Kibble-Zurek mechanism with variable
results. While the numerical studies claimed that they successfully
verified the predictions, the conclusions are less clear in the
experimental works that studied vortex formation in superfluid $^4$He
and $^3$He with null results in the former~\cite{no} and agreement
with the KZ prediction in the latter~\cite{yes}. See \cite{newexp,kirtley,kibble-exp}
for discussions of some recent experimental results disagreeing with the KZ prediction.

In the following we revisit the KZ scaling analysis. We focus on the
dynamics of classical systems coupled to an environment, a setting in
which the dynamics are stochastic and the energy, directly linked to
the number of topological defects, is not conserved. We use the
temperature of the thermal bath as the control parameter driving the
second order phase transition and a linear cooling rate
$T(t) = T_c ( 1 - t/\tau_Q)$.
These choices are made to keep the discussion simple; extensions to
more complicated protocols are straightforward and will be partially
addressed later.  Moreover, we restrict to systems with a unique
equilibrium correlation length and a single dynamic
counterpart, the typical growing length. We solely deal with problems
with power-law scaling laws, that link {\it e.g.}  the correlation
length to the distance from criticality, and length-scales to
time-scales~\cite{foot3}. These restrictions exclude from the analysis
complex systems with several competing lengths and problems with
quenched disorder.  Henceforth we focus on the growth law of the size
of the correlated regions, $R(t)$.  This
will allow us to discuss systems characterized by topological defects
as well as those that are not from the same point of view. We shall explain below how the density
of topological defects can be obtained from $R(t)$.

Let us start our analysis with some simple remarks. First,  although the initial 
adiabatic regime and the departure from it are expected, the existence of the 
impuse regime is questionable. During this regime, which would take place between $-\hat t$ and
$\hat t$, the system is supposed not to
evolve~\cite{original-Kibble,original-Zurek,report-Zurek,LesHouches}.
However, it is well-known that after a rapid quench into the critical 
region any system undergoes {\it critical coarsening} described by
the growth of a {\it typical linear length-scale} for correlated
regions, $R(\Delta t) \sim \Delta t^{1/z_{eq}}$ where $\Delta t$ is
the time spent in the critical region  and $z_{eq}$ is the exponent that
links the equilibrium relaxation time and correlation length close to
criticality~\cite{Beate,Calabrese-Gambassi}.
Above criticality, the major difference between slow and rapid
quenches is in the extension of the adiabatic regime.  The slower the
quench or the annealing, the closer the system gets to the critical point in
equilibrium.  However, also for very slow annealing, the system
eventually departs from the adiabatic evolution and has to undergo
critical coarsening.

Our second remark is that once getting across the critical point, when the running temperature
$T(t)$ is far enough from $T_c$, the dynamics crosses over
to standard coarsening.  In order to get a better insight into this
process, let us recall that an infinitely rapid quench to a
temperature $T<T_c$ leads to a growth law $R\simeq \lambda(T) \Delta
t^{1/z_d}$ ~\cite{Alan}. Now $z_d$ is the {\it dynamic exponent} that,
quite generally, is different from $z_{eq}$ and depends on the dynamic
rules. The
prefactor vanishes at $T_c$ and is characterized by a singular power
law $\lambda(T) \simeq
|T-T_c|^{\nu(-1+z_{eq}/z_d)}=\xi_{eq}^{1-z_{eq}/z_d}$~\cite{Sicilia-PRE}.
If the annealing rate is finite, one naturally expects the growth law at
long times and far from the critical point to be $R\simeq
\lambda(T(t)) \Delta t^{1/z_d}$. The reason is that the dynamical
process renormalizing the value of $\lambda(T(t))$ should be finite
and, hence, evolve on a much faster timescale than the coarsening one which
instead is of the order of the age of the system and diverges with
$t$.\\
\begin{figure}
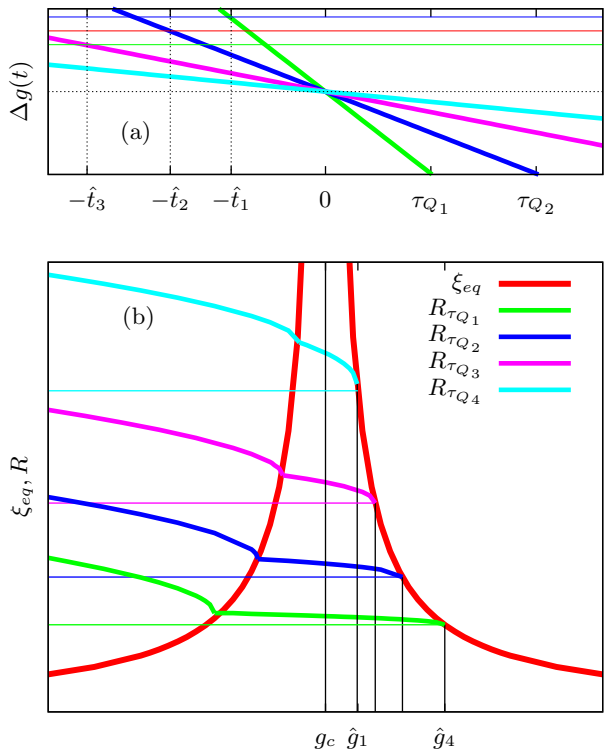

\input{fig-kz-coolingrate.pslatex}
\input{fig-kz-sketch-length-stell.pslatex}
\caption{(Colour online.) (a) The
  control parameter, $\Delta g(t)=[g(t)-g_c]/g_c$, for different
  cooling rates. The crossover between adiabatic and out of
  equilibrium dynamics are signaled as $-\hat t_i$ for ${\tau_Q}_i <
  {\tau_Q}_{i+1}$.  (b) Sketch of the control parameter dependence of
  the equilibrium correlation length (thick red line)  and the dynamic growing
  length $R$ for four linear cooling rate proceedures with ${\tau_Q}_i
  < {\tau_Q}_{i+1}$. Values of the control parameter at which the
  dependence changes from adiabatic to critical are shown as $\hat
  g_{i}$ (for simplicity we plot them as singular points in the evolution of $R$. 
  In reality they just correspond to cross-overs). For comparison the assumption of constancy
  during the impulse and subcritical regimes are shown with thin
  horizontal lines.}
\end{figure}

We now endeavor to connect the dots and propose a
general scenario for infinitely slow annealing. Our main conjecture, that is motivated
by the previous discussion and the fact that the system stays for a very long time in the vicinity 
of the critical point, involves the growth of the length-scale $R(t)$:
\begin{equation}
R(t) \simeq \xi_{eq}(T(t)) \ f\left[\frac{t}{\tau_{eq}(T(t))}\right]
\; .
\label{eq:scaling-hypothesis}
\end{equation} 
This asymptotic form encompasses equilibrium above the critical
point, $x\equiv t/\tau_{eq}(T(t))\ll -1$, critical coarsening,
$x\propto O(1)$, and the cross-over to standard coarsening $x\gg 1$.
The limits of $f(x)$ are obtained by requiring to find the expected
adiabatic behavior and standard coarsening on the two extremes,
\begin{eqnarray}
R(t) \simeq \left\{
\begin{array}{ll}
\xi_{eq}(T(t)) & \qquad t  \ll -\tau_{eq}(T(t)) \; , \\
\left[\xi_{eq}(T(t))\right]^{1-\frac{z_{eq}}{z_d}} \ t^{\frac{1}{z_d}} 
& \qquad t \gg \tau_{eq}(T(t)) 
\; .
\end{array}
\right.
\label{eq:dynamic-scaling}
\end{eqnarray}
This imposes that $f(x)$ be a constant for $x\ll -1$ and proportional
to $ x^{1/z_d}$ for $x\gg 1$. We expect the scaling function $f(x)$ to
be universal since it describes evolution on diverging time and length
scales close to the critical point. A sketch of $\xi_{eq}$ and $R$ is
shown in Fig. 1.
Our scaling assumption applies to coarsening with and without
topological defects.  In the former case, the {\it decaying typical
  number of topological defects}, $N(t) \simeq R(t)^{-d}$,
reads for $t\gtrsim \tau_Q$
\begin{equation}
N(t) \simeq \tau_Q^{\frac{d\nu}{z_d} (z_{eq}-z_d)} 
t^{-\frac{d}{z_d} [1+\nu (z_{eq}-z_d)]}
\; . 
\label{eq:our-prediction}
\end{equation}
Note that a qualitatively similar dependence on $t$ and $\tau_Q$ was
found numerically in~\cite{williams} in a system with
vortex-antivortex pairs.  Evaluating the above expression at $t=\hat
t=\tau_Q^{\nu z_{eq}/(1+\nu z_{eq})}$ we recover KZ's result,
eq~(\ref{eq:KZ-prediction}). Ours, however, is more general since it
applies to any time $t$ and it allows one to describe all the slow
annealing evolution. For example, $N(t)\propto
\tau^{-d/z_d}_Q$ on times of the order of the inverse annealing rate, $t\simeq \tau_Q$, thus showing that a
substantial decrease takes place after $\hat t$. In comparison with
the KZ mechanism, our arguments allow one to understand why $R(\hat
t)\simeq f R(-\hat t)$ with a factor $f$ that can be as large as $10$ 
in some cases \cite{Laguna-Zurek}.  This was somewhat mysterious in the KZ scenario where
defects are frozen out in the impulse regime.  We understand the
reduction in the number of topological defects 
as due to critical coarsening. Taking this phenomenon into
account is crucial for more general annealing protocols, e.g.
$T(t)=\theta (-t)T_c(1-t/\tau_Q^{(1)})+\theta
(t)T_c(1-t/\tau_Q^{(2)})$.  For a large ratio
$\tau_Q^{(2)}/\tau_Q^{(1)}$ the system spends a long time in the
critical region and $R(t)$ evolves during the critical coarsening from $[\tau_Q^{(1)}]^{\nu/(1+\nu
  z_{eq})}$ to $[\tau_Q^{(2)}]^{\nu/(1+\nu z_{eq})}$.

In the following we provide numerical evidence for the conclusions
outlined above by presenting results of a Monte Carlo simulation (using the heat bath 
algorithm with random sequential updates) of
the $2d$ Ising model (IM) on square and triangular lattices. In particular,
we check eqs.~(\ref{eq:scaling-hypothesis})-(\ref{eq:our-prediction})
and the universality of the scaling function $f(x)$. 
We equilibrate the system at $T_0=2T_c$ and
we use a linear cooling rate  that takes
the temperature of the bath from $T_0$ at
$t=-\tau_Q$ to $T=0$ at $t=\tau_Q$.
We find that the system undergoes an adiabatic
evolution until it falls out of equilibrium close to $T_c$. The
typical correlation length, $R(t)$, is extracted from the analysis of
the space-time correlation $C(r,t) \equiv \langle s(\vec x) s(\vec
x+\vec r) \rangle \simeq g(r/R(t))$ with the average taken over
$100$ initial conditions and 
noise realizations. 
We use various ways to determine $R$ and verify that they yield equivalent
results. Two of them are $C(R(t),t)=1/2$ and $R(t) = \int d^2r \
r^\zeta C(r,t)/\int d^2r \ r^{\zeta-1} C(r,t)$ with $\zeta$ a parameter
that is chosen for convenience, namely to weigth differently shorter
or longer distances.
\begin{figure}[h]
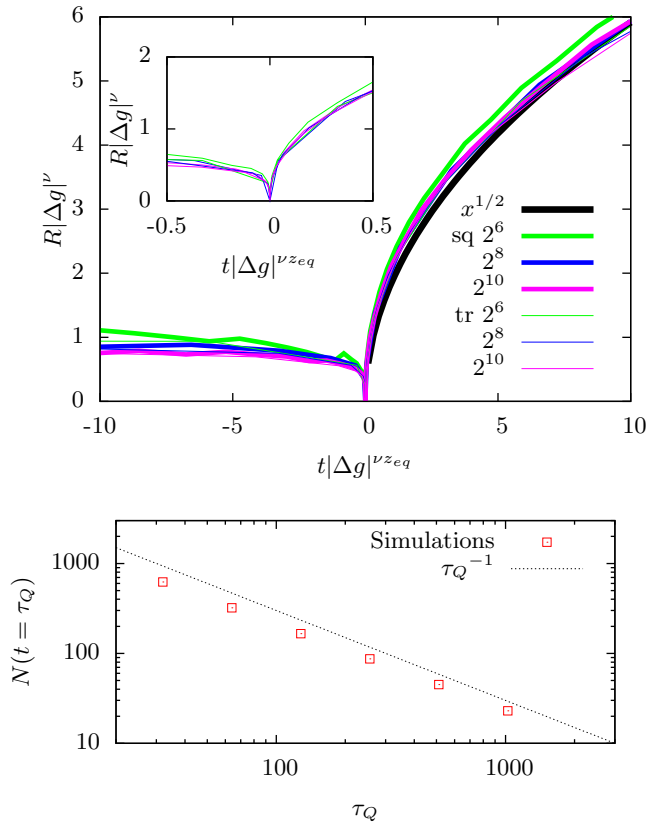

\input{KZ_scaling.pstex}
\input{N_at_zeroT.pstex}
\caption{(Colour online.) Top panel: Test of the dynamic scaling
  hypothesis~(\ref{eq:scaling-hypothesis}) and the limits
  (\ref{eq:dynamic-scaling}) in the $2d$IM on a triangular and a
  square lattice, annealed at different cooling rates given in the
  key. A zoom on the critical region $|x|\stackrel{<}{\sim} 1$ is
  shown in the inset.  The exponents are $\nu=1$, $z_{eq}=2.17$ and
  $z_d=2$. Bottom panel: Number of deffects at $t=t_Q$ for different
  cooling rates. Points represent the numerical data while the line
  corresponds to the prediction $N(\tau_Q)={\tau_Q}^{-d/z_d}$}
\label{fig:test-scaling}
\end{figure}
In the top panel of Fig.~\ref{fig:test-scaling} we test the scaling hypothesis,
eq.~(\ref{eq:scaling-hypothesis}), and the limits of the scaling
function $f(x)$, eq.~(\ref{eq:dynamic-scaling}). For both the square
and triangular lattices we find very good agreement between numerical
data and theoretical expectation. The square root growth at positive
times demonstrates that standard coarsening cannot be ignored. 
The scaling collapse improves, as expected, restricting the range of $|x|$.  
A zoom on the small $|x|$ region is shown in the inset.
Moreover, we find that the scaling functions for square and
triangular lattices coincide within numerical accuracy, confirming
the universality of $f(x)$. The bottom panel of Fig~\ref{fig:test-scaling} displays $N(\tau_Q)$ 
and confirms the $\tau_Q^{-d/z_d}$ decay - the power is shown 
as a guide-to-the-eye next to the data. 

A further test is provided by the
 analytic solution to the evolution
of the $O(N)$ model in the large $N$ limit for a very slow annealing.  
This is a $\lambda \phi^4$ field theory in which the order
parameter is upgraded to an $N$-dimensional vector and the fourth
order term in the double-well potential is conveniently normalized to
allow for an $N\to\infty$ limit in which the model becomes solvable
but still non-trivial \cite{Alan}.  Note that although there are no
topological defects, since the large $N$ limit is taken at fixed
dimension, $N\gg d$, the dynamics are still characterized by a growing
correlation length $R(t)$.  The analysis of a finite rate quench is a
simple generalization of the treatment of infinite rate ones (see,
{\it e.g.}~\cite{Alan}). We find that the
scaling~(\ref{eq:dynamic-scaling}) holds with $\nu=1/2$ and
$z_d=z_{eq}=2$ in all $d>2$. Due to the coincidence of the $z$
exponents the prefactor in the bottom expression of eq.~(\ref{eq:dynamic-scaling}) equals one
and the dependence on $\tau_Q$ disappears. 

As a summary we analyzed the dynamic evolution induced by annealing
with rate $1/\tau_Q$ ($\tau_Q\rightarrow \infty$) in pure systems
characterized by conventional dynamic scaling and standard low
temperature coarsening.  We obtained a complete picture of the
dynamics which is characterized by three regimes:
adiabatic, critical coarsening and standard coarsening.  Using scaling
arguments we found the growth law of the correlation length during the
annealing and its $\tau_Q$ scaling dependence. The cross-over between
adiabatic and coarsening regimes is governed by a universal scaling
function. We tested our findings with numerical simulations of the
$2d$ Ising model and a large $N$ analysis of the $O(N)$ model in
$d> 2$. Our results generalize the KZ mechanism and,
at the same time, show its limitation. In particular we find that the
defect dynamics are not frozen in the so-called impulse regime, as it
can be found by using more general annealing protocols than a linear ramp
in temperature.

Physical situations in which understanding the evolution during a slow
annealing is important and which we plan to study in the future are
disordered and quantum systems. Several studies have dealt with the
former, see e.g. \cite{fisherhuse,yoshino,leticiajp}. The latter have only
recently received attention in connection with quantum quenches and annealing in
cold atoms. In these cases the conditions are different from the ones
analyzed in this paper since {\it isolated} systems in which a
coupling is slowly changed through a quantum critical point are
usually considered. The absence of the thermal bath may change
drastically the physics.  The KZ mechanism has been argued to apply
{\it mutatis mutandis} to the isolated quantum
case as well~\cite{Zurek-Zoller,Polkovnikov}. This has 
been verified in some integrable cases~\cite{Dziarmaga}. 

We close with a note on an exact study of the 
cooling rate effects in the relaxation 
of the classical Ising chain with Glauber 
dynamics by P. Krapivsky that shows qualitative but not 
quantitative agreement with the KZ mechanism~\cite{Krapivsky}.

This work was financially supported by ANR-BLAN-0346 (FAMOUS).  We
thank C. Kollath, C. Godreche, A. Gambassi, D. A. Huse, A. Jelic,
P. Krapivsky, J. Kurchan, G. Lozano and A. Silva for very useful
discussions.


\begin{references}

\bibitem{Alan} A. J. Bray, Adv.\ Phys.\ {\bf 43}, 357 (1994).

\bibitem{Cugliandoloreview}
J.-P. Bouchaud, L. F. Cugliandolo, J. Kurchan, M. M\'ezard 
in ``Spin-glasses and random fields'', A. P. Young Ed. (World Scientific).

\bibitem{footnote0}
The picture
  for cases characterized by the existence of topological defects is
  slightly different, see e.g.~\cite{Alan}.

\bibitem{fisherhuse}
D.A. Huse and D.S. Fisher, Phys. Rev. Lett. {\bf 57} 2203 (1986).

\bibitem{yoshino}
H. Yoshino, K. Hukushima, H. Takayama
Phys. Rev. B 66, 064431 (2002).

\bibitem{leticiajp}
F. Alberici-Kious, J.-P. Bouchaud, L. F. Cugliandolo, P. Doussineau, A. Levelut,  Phys. Rev. B {\bf 62}, 14766 (2000).

\bibitem{original-Kibble} T. W. B. Kibble, J. Phys. A {\bf 9}, 1387 (1976).

\bibitem{original-Zurek} W. H. Zurek, Nature (London) {\bf 317}, 505 (1985).

\bibitem{report-Zurek}
W. H. Zurek, Phys. Rep. {\bf 276} 177 (1996).

\bibitem{LesHouches} W. H. Zurek, L. M. A. Bettencourt, 
J. Dziarmaga, and N. D. Antunes, {\it Shards of broken symmetry} in
Topological defects and the non-equilibrium dynamcis of symmetry breaking phase
transitions,  Vol 549, Y. M. Bunkov and H. Godfrin eds. 
 (Kluwer Academic Publishers, 1999). 

\bibitem{Zurek-Zoller}
W. H. Zurek,  U. Dorner, P.  Zoller, Phys. Rev. Lett.
{\bf 95}, 105701 (2005). 

\bibitem{Polkovnikov} A. Polkovnikov, 
Phys. Rev. B {\bf 72},  161201 (2005). 

\bibitem{Dziarmaga} J. Dziarmaga, Phys. Rev. Lett. {\bf 95}, 245701 (2005).

\bibitem{Hohenberg-Halperin} 
P. C. Hohenberg and B. Halperin, Rev. Mod. Phys. {\bf 49}, 435 (1977).

\bibitem{Laguna-Zurek} P. Laguna and W. H. Zurek, Phys. Rev. Lett. {\bf 78}, 25
19 (1997); Phys. Rev. D {\bf 58}, 5021 (1998). 

\bibitem{Yates-Zurek} A. Yates and W. H. Zurek, Phys. Rev. Lett. {\bf 80}, 5477 (1998). 

\bibitem{Bettencourt-Zurek} N. D. Antunes, L. M. A. Bettencourt, W. H. Zurek, 
Phys. Rev. Lett. {\bf 82}, 2824 (1999). 

\bibitem{Stephens} G. J. Stephens, E. A. Calzetta, B. L. Hu, and S. A. Ramsey, 
Phys. Rev. D {\bf 59}, 045009 (1999). 

\bibitem{Suzuki} 
S. Suzuki, {\it Cooling dynamics of pure and random Ising chains}, 
arXiv:0807.2933.

\bibitem{no} M. E. Dodd, P. C. Henrym\, N. S. Lawson,
P. V. E. McClintock, and C. D. H. Williams, Phys. Rev. Lett. {\bf 81},
3703 (1998).

\bibitem{yes} V. M. H. Ruutu, V. B. Eltsov, A. J. Gill,
T. W. B. Kibble, M. Krusius, Y. G. Makhlin, B. Placais, G. E. Volovik,
W. Xu, Nature {\bf 382}, 334 (1996).  C. B\"auerle, Y. M. Bunkov,
S. N. Fisher, H. Godfrin, G. R. Pickett, Nature {\bf 382}, 332 (1996).
{\it The Grenoble cosmological experiment: the Kibble-Zurek scenario
in superfluid $^3$He}, C. Bauerle, Yu. M. Bunkov, S. Fisher, and
H. Godfrin, in Topological defects and the non-equilibrium dynamcis of
symmetry breaking phase transitions, Vol 549, Y. M. Bunkov and
H. Godfrin eds.  (Kluwer Academic Publishers, 1999).

\bibitem{kibble-exp} T. Kibble, Physics Today, {\bf 60}, No. 9, 47 (2007).

\bibitem{kirtley}
J. R. Kirtley and F. Tafuri, Physics {\bf 2}, 92 (2009).

\bibitem{newexp}
R. Monaco, J. Mygind, R. J. Rivers and V. P. Koshelets, Phys. Rev. B, {\bf 80},
180501 (2009).

\bibitem{foot3} Some cases in which the dynamic scaling hypothesis
fails due to more than one growing length exist~\cite{Alan}.

\bibitem{Beate} H. K. Janssen, B. Schaub, and B. Schmittmann,
  Z. Phys. B {\bf 73}, 539 (1989).  B. Schmittmann and R. K. P. Zia in
  {\it Phase transitions and critical phenomena} C. Domb and
  J. L. Lebowitz Vol. {\bf 17} (Academic Press, N. Y.)  1995.

\bibitem{Calabrese-Gambassi} 
P. Calabrese and A. Gambassi, J. Phys. A {\bf 38}, R133 (2005). 
U. C. Tauber 
Lecture Notes in Physics {\bf 716}, 295 (2007). 

\bibitem{Sicilia-PRE}
A. Sicilia, J. J. Arenzon, A. J. Bray, and L. F. Cuglandolo, 
Phys. Rev. E {\bf 76}, 061116 (2007).

\bibitem{williams} 
H.-C. Chu and G. A. Williams
Phys. Rev. Lett. {\bf 86}, 2585  (2001).

\bibitem{Krapivsky}
P. Krapivsky, arXiv:cond-mat (2010). 

\end{references}
\end{document}